\documentclass[preprint,showpacs,preprintnumbers,amsmath,amssymb]{revtex4}

\usepackage{graphicx}
\usepackage{dcolumn}
\usepackage{bm}

\begin{document}

\preprint{APS/123-QED}

\title{Time Quantified Monte Carlo Algorithm for Interacting Spin Array Micromagnetic Dynamics}

\author{X. Z. Cheng}
\author{M. B. A. Jalil}%
\affiliation{%
Department of Electrical and Computer Engineering, National
University of Singapore, 4 Engineering Drive 3, 117576, Singapore
}%

\author{Hwee Kuan Lee}
\affiliation{ Data Storage Institute, 5 Engineering Drive 1, DSI
Building, 117608, Singapore
}%

\date{\today}

\begin{abstract}
In this paper, we reexamine the validity of using time quantified
Monte Carlo (TQMC) method [Phys. Rev. Lett. 84, 163 (2000); Phys.
Rev. Lett. 96, 067208 (2006)] in simulating the stochastic dynamics
of interacting magnetic nanoparticles. The Fokker-Planck
coefficients corresponding to both TQMC and Langevin dynamical
equation (Landau-Lifshitz-Gilbert, LLG) are derived and compared in
the presence of interparticle interactions. The time quantification
factor is obtained and justified. Numerical verification is shown by
using TQMC and Langevin methods in analyzing spin-wave dispersion in
a linear array of magnetic nanoparticles.
\end{abstract}

\maketitle

\section{\label{sec:level1}Introduction}
The TQMC method is found to be a powerful simulation technique in
modeling magnetization reversal dynamics of magnetic nanoparticles
\cite{nowak,hinzke,chubykaloTQMC,chengprb,chengprl}. It is found
that simulation with the TQMC method is considerably more efficient
than the conventional method of modeling magnetization dynamics
based on time-step integration of the stochastic LLG equation,
especially in the case of high damping limits \cite{chubykaloTQMC}.
The attraction of TQMC also lies on the fact that it establishes an
analytical connection between the two stochastic simulation schemes,
Monte Carlo (MC) and Langevin dynamics, which were previously
thought to have different theoretical bases. Such analytical
connection provides alternative techniques to both stochastic
models, e.g. solving a stochastic differential equation using
advanced Monte Carlo techniques to calculate the long-time reversal
\cite{chengprb,chubykaloKINETIC}.

The validity of using TQMC to simulate an isolated single domain
particle is first demonstrated by Nowak \emph{et al.} \cite{nowak}
and later rigorously proved by us in Ref.~\cite{chengprl} using the
Fokker-Planck equation as a bridge between MC and Langevin methods.
In the case of interacting spin arrays, the validity of TQMC has not
been analytically proved although it has been numerically shown
\cite{hinzke,chengprl}. It thus comes the necessity to establish the
proof for the case of interacting spin systems, since in practical
applications the discrete spins or moments (in the form of magnetic
nanoparticles) are usually closely packed together, and hence are
strongly coupled to one another. It is also important to show
explicitly whether the analytical equivalence between the TQMC and
the stochastic LLG equation and the time quantification factor, are
dependent in any way on the nature (e.g. magnetostatic or exchange)
or strength of the coupling interactions.

In this paper, we provide a rigorous proof for this case, based on
the technique presented in our earlier works \cite{chengprl}. We
further demonstrate the generality of the TQMC method, by
implementing it in two different contexts, i.e. in time-evolution
and reversal studies of a square array of spins, and in analyzing
the spin wave dispersion in a linear spin chain.

\section{\label{sec:level2}Model}
The physical model under consideration is a spin array (which
represent an array of magnetic nanoparticles), whose spin
configuration is represented as $\{\mathbf{s}\} =
\{\cdots,\hat{\mathbf{s}}_{i},\cdots\}$, where $\mathbf{s} =
\mathbf{M}/M_s$ is a normalized unit vector representing the
magnetic moment of the spin and $i$ refers to the $i^{\mathrm{th}}$
spin in the vector list of length $N$.

The micromagnetic dynamics of the spin array is traditionally
described by the Landau-Lifshitz-Gilbert (LLG) equation:
\begin{equation} \label{eq:llg}
 \frac{d}{dt} \{\mathbf{s}\}
  = -\frac{\gamma_0H_k}{1+\alpha^2} \{\mathbf{s}\} \times
   \left(
     \{\mathbf{h}\}+\alpha\cdot\{\mathbf{s}\} \times \{\mathbf{h}\}
   \right)
\end{equation}
where $\alpha$ and $\gamma_0$ are the damping constant and the
gyromagnetic constant respectively, $\{\mathbf{h}\} =
\frac{1}{2K_uV} \nabla_{\{\mathbf{s}\}} E$ is the effective field
which is normalized with respect to the anisotropy field
$H_k=2K_u/\mu_0 M_s$, where $K_u$ is the anisotropy constant and
$\mu_0$ is the magnetic permeability. $E = E(\{\mathbf{s}\})$ is the
total energy of the system which consists of the typical
contributions in a micromagnetic system, e.g., Zeeman term,
anisotropy term, magnetostatic term and exchange coupling term. The
operator $\{\mathbf{s}\} \times \{\mathbf{h}\} =
\{\cdots,\mathbf{s}_i \times \mathbf{h}_i,\cdots\}$ is understood.
To represent the thermal fluctuation, white noise-like stochastic
thermal fields are added to the effective field according to Brown
\cite{brown63}.

Alternatively, Random walk Monte Carlo (MC) algorithm can also be
used in simulating the magnetization reversal dynamics \cite{nowak}.
At each Monte Carlo step, one of the $N$ spin sites is randomly
selected to undergo a trial move, in which a random displacement
lying within a sphere of radius $R$ ($R \ll 1$) is added into the
original magnetic moment and the resulting vector is then
renormalized. The magnetic moment changes according to a heat bath
acceptance rate as $A(\Delta E) = 1 / (1 + \exp(\beta \Delta E))$.
Here $\Delta E$ is the energy change within the random walk step and
$\beta=(k_BT)^{-1}$, $k_B$ is the Boltzmann constant and $T$ is the
temperature in Kelvin.

\section{\label{sec:level3}Fokker-Planck Equations}

To link the MC scheme with the stochastic LLG equation, we shall
derive the respective Fokker-Planck (FP) coefficients corresponding
to the LLG equation and the random walk MC \cite{chengprl}. The
general Fokker-Planck equation (FPE) for a spin array in a spherical
coordinates is given as
\begin{eqnarray}
  \frac{d}{dt}P(\{\theta\},\{\varphi\},t)
    &=&
    - \sum_i \frac{\partial}{\partial\theta_i} \left(A_{\theta_i} \cdot P\right)
    - \sum_i \frac{\partial}{\partial\varphi_i} \left(A_{\varphi_i}\cdot P\right)
    +\frac{1}{2}\sum_{i,j}\frac{\partial^2}{\partial\theta_i\partial\theta_j}\left(B_{\theta_i\theta_j}\cdot P\right)
  \nonumber\\
    & &
    +\frac{1}{2}\sum_{i,j}\frac{\partial^2}{\partial\varphi_i\partial\varphi_j}\left(B_{\varphi_i\varphi_j}\cdot P\right)
    + \frac{1}{2} \sum_{i,j} \frac{\partial^2}{\partial\theta_i\partial\varphi_j} \left((B_{\theta_i\varphi_j}+B_{\varphi_j\theta_i})\cdot P\right)
\label{eq:fpe}
\end{eqnarray}
where drift terms $A_x$ and diffusion terms $B_{xy}$ ($x =
\{\theta_i, \varphi_i\}$, $y = \{\theta_j, \varphi_j\}$) are defined
as the ensemble mean of an infinitesimal change of $x$ and $y$ with
respect to time \cite{risken}. By giving the detailed derivation in
the appendix, we obtained the Fokker-Planck coefficients for LLG:
\begin{eqnarray}
  \label{eq:ldfpe}
    A^{LLG}_{\theta_i} &=& -h'\frac{\partial E}{\partial\theta_i}
                           -g'\frac{1}{\sin\theta_i}\frac{\partial E}{\partial\varphi_i}
                           +k'\cot\theta_i
  \nonumber\\
    A^{LLG}_{\varphi_i}&=& g'\frac{1}{\sin\theta_i}\frac{\partial E}{\partial\theta_i}
                           -h'\frac{1}{\sin^2\theta_i}\frac{\partial E}{\partial\varphi_i}
  \nonumber\\
    B^{LLG}_{\theta_i\theta_j}&=&2k'\cdot\delta_{ij}
  \\
    B^{LLG}_{\varphi_i\varphi_j}&=&\frac{1}{\sin^2\theta_i}2k'\cdot\delta_{ij}
  \nonumber\\
    B^{LLG}_{\theta_i\varphi_j}&=& B^{LLG}_{\varphi_j\theta_i} = 0
  \nonumber
\end{eqnarray}
\\
as well as for the TQMC:
\begin{eqnarray}
 \label{eq:mcfpe}
    A_{\theta_i}^{MC}&=& N^{-1}\frac{R^2}{20}\left(\cot\theta_i-\beta\frac{\partial E}{\partial\theta_i}\right) 
  \nonumber\\
    A_{\varphi_i}^{MC}&=& -N^{-1}\frac{1}{\sin^2\theta_i}\frac{R^2}{20}\beta\frac{\partial E}{\partial\varphi_i} 
  \nonumber\\
    B_{\theta_i\theta_j}^{MC}&=& N^{-1}\frac{R^2}{10}\cdot\delta_{ij} 
  \\
    B_{\varphi_i\varphi_j}^{MC}&=& N^{-1}\frac{1}{\sin^2\theta_i}\frac{R^2}{10}\cdot\delta_{ij}
  \nonumber\\
    B_{\theta_i\varphi_j}^{MC}&=& B_{\varphi_j\theta_i}^{MC} = 0
  \nonumber
\end{eqnarray}
where in Eq.~(\ref{eq:ldfpe}),
$h'=\frac{\alpha\gamma_0}{\mu_{0}VM_{s}(1+\alpha^{2})}$,
$g'=h'/\alpha$, $k'=h'/\beta$.

\section{\label{sec:level4}Mapping MC to LLG}
In the high damping limit where the damping constant $\alpha$ is
large, so that $g' = h'/\alpha \to 0$, a term-wise equivalence can
be established between the FPE coefficients in Eqs. (\ref{eq:ldfpe})
and (\ref{eq:mcfpe}), corresponding to the LLG and MC methods, if:
\begin{equation}
\label{eq:timeq}
    R^2 \Delta \tau_{\mathrm{\scriptscriptstyle MC}}=
        \frac{20 \alpha}{1+\alpha^2} \frac{\gamma_0}{\beta \mu_{0} V M_s} \Delta t_{\mathrm{\scriptscriptstyle LLG}}.
\end{equation}

Eq.~(\ref{eq:timeq}), in which $\Delta
\tau_{\mathrm{\scriptscriptstyle MC}}$ is calibrated in MCS/site
(one Monte Carlo step for each site on the average), is the time
quantification factor for the TQMC method in interacting spin
arrays. The time quantification factor is found to be the same as
the one in Ref. \cite{nowak} for an isolated single particle case,
and is thus consistent with previous numerical convergence observed
in Refs. \cite{hinzke,chengprl}.

For the low damping limit where precessional motion becomes
significant, one may wish to use the precessional (hybrid)
Metropolis Monte Carlo algorithm \cite{chengprl}. We confirm that,
by using the same derivation techniques, one is able to prove the
validity of including the precessional move in the MC algorithm in
simulating the micromagnetic properties of an interacting spin
array.

\section{\label{sec:level5}Results and Discussion}

\begin{figure}
\includegraphics[width=0.5\textwidth, bb={10 10 280 210}]
{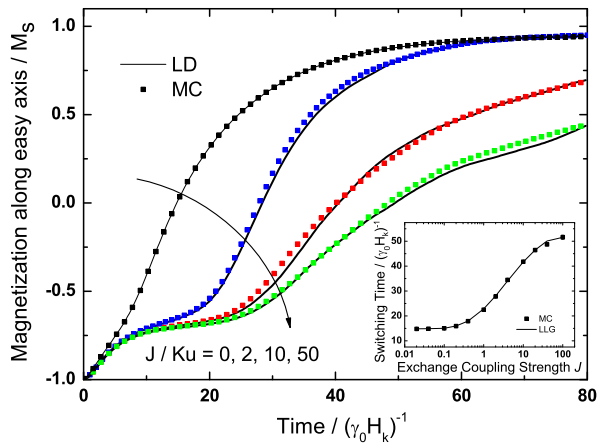}
\caption{\label{fig:Jex} (color online) The time evolution behavior
of the magnetization reversal in a spin array system. The following
simulation parameters are assumed: lattice size of $10\times10$,
periodical boundary condition, thermal condition $K_uV/k_BT = 25$,
damping constant $\alpha = 1.0$ and external field $h = 0.5$ applied
on an angle $\theta = \pi/4$ with respect to the easy axes. The
exchange coupling strength $J$ is the adjustable variable. To
guarantee the simulation accuracy, the time interval $\Delta t$ for
the LLG integration changes with $J$ as $\Delta t =
0.01/(1+h+J/K_uV)$ \cite{dimitrov}, while the trial move step size
$R$ in the MC simulation is chosen to reflect the $\Delta t$ in one
MCS. Error bars are smaller than the symbol size.}
\end{figure}

The equivalence between the MC method and LLG, which is expressed by
Eq.~(\ref{eq:timeq}), provides the theoretical justification for the
use of MC method as an alternative to the LLG equation in
micromagnetic studies. The equivalence which has been established is
very general because no explicit form of the Hamiltonian is used in
the derivation.  This implies that the validity of the equivalence
is independent of many physical and simulation parameters.  For
illustration, we test the validity of the TQMC method for a simple
$10\times10$ spin array which is subject to a varying exchange
coupling strength $J$. As shown in Fig.~(\ref{fig:Jex}), the time
evolution behavior of the (asymmetric) magnetization reversal is
simulated for different values of $J$. We find good convergence
between the simulated results from both LLG and MC schemes, even
when the switching mechanism of the spin array changes from the
independent reversal (small $J$) to the nucleation-driven reversal
(large $J$). We also confirm that the mapping between MC and LLG
time steps as expressed in Eq.~(\ref{eq:timeq}), is also independent
of other simulation and physical parameters, e.g. the chosen
boundary condition (periodic / free), the lattice size, and the
nature of the coupling (magnetostatic / exchange).

Next, we show that the equivalence between the MC method and LLG
enables the MC method to be utilized in most of the situations where
LLG applies, and beyond the above time-evolution simulation. As an
example, we consider the dispersion relation for the primary spin
wave mode of a one-dimensional spin chain. This example is chosen
because it tests the capability of precessional TQMC method to
simulate both spatial and time correlation of the spin-wave
dynamics. By comparison, conventional MC methods are more suited for
equilibrium or steady-state studies rather than time correlation
dynamics.

\begin{figure}
\includegraphics[width=0.5\textwidth, bb={10 10 280 210}]
{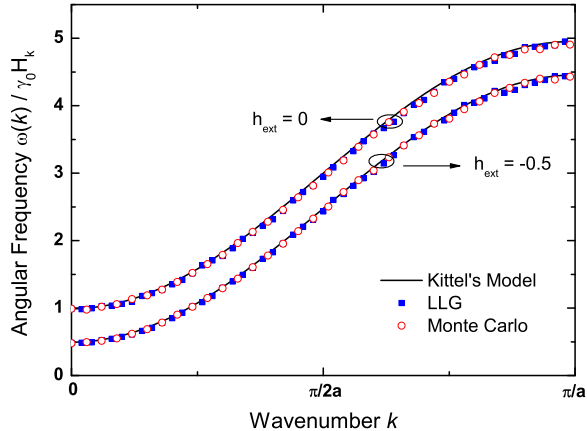}
\caption{\label{fig:spinwave} (color online) Dispersion relation for
the simulated spin wave mode. Simulation parameters are: chain
length $N = 200$, free boundary condition, thermal condition
$K_uV/k_BT = 50$, exchange coupling strength $J/2K_uV=1$ and damping
constant $\alpha = 0.1$. Kittel's model refers to the theoretical
dispersion relation of Eq.~(\ref{eq:dispersion}).}
\end{figure}

The Hamiltonian of the spin chain system is set to be:
\begin{equation}\label{hamiltonian}
  \mathcal{H} = \sum_{i} \left(
    - J \sum^{\{i\}}_{j < i} \hat{\mathbf{s}}_i \cdot \hat{\mathbf{s}}_j
    - K_u V (\hat{\mathbf{s}}_i \cdot \hat{\mathbf{k}}_n)^2
    - \mu_0 M_s V \cdot \hat{\mathbf{s}}_i \cdot \mathbf{H}_{\mathrm{ext}}
  \right)
\end{equation}
where $\{i\}$ represents the neighboring spins of the
$i^{\mathrm{th}}$ spin, $J$ is the coupling strength,
$\mathbf{H}_{\mathrm{ext}}$ is the applied field and
$\hat{\mathbf{k}}_n$ refers to the unit vector along the easy axis.
Magnetostatic coupling was not included in this test. The dispersion
relation for the one-dimensional spin wave mode has been
theoretically studied \cite{kittel} and is given by:
\begin{equation}\label{eq:dispersion}
    \omega(k) = \frac{\gamma_0 H_k}{1 + \alpha^2}[1 + h_{\mathrm{ext}}+ 4(J/2K_uV)\sin^2(ka/2) ]
\end{equation}
where $h_{\mathrm{ext}} = H_{\mathrm{ext}}/H_k$ and $a$ is the
lattice constant. The calculations were done using the computational
techniques of Refs.~\cite{Chantrell,chubykaloWAVE}. Spins were
initially aligned along the $z$ direction, in parallel with both the
easy axes and applied fields. Stochastic simulation was performed on
this initial configuration for 100 $(\gamma_0 H_k)^{-1}$, in order
to achieve the quasi-equilibrium state. Space and time Fourier
transforms were then performed on the off-axis components. From the
resulting spin wave spectra, the peak frequency $\omega$ determined
for a range of wavevector $k$. The resulting dispersion relation in
Fig.~(\ref{fig:spinwave}) shows a very good convergence between the
simulated results (calculated from both LLG and MC) and the
theoretical prediction of Eq.~(\ref{eq:dispersion}).

However, there do exist limitations to the TQMC method. The TQMC
method cannot be easily extended for some advanced micromagnetic
simulations, which include e.g. the spin torque effect, as compared
to the stochastic LLG model. In addition, even though we found the
TQMC algorithm to be typically 2--5 times faster than the LLG-based
simulation with an equivalent time step, it is still too inefficient
to model long-time magnetization reversal of up to, say, 1 second.
Nevertheless, our analysis has raised the possibility of developing
advanced time-quanti¡¥able Monte Carlo methods, based on e.g. the
N-fold way Monte Carlo algorithm \cite{novotny} and kinetic Monte
Carlo method \cite{chubykaloKINETIC}, for micromagnetic studies.

\section{\label{sec:level6}Conclusion}
We have derived the time quantification factor which relate the
time-scales of TQMC method and Langevin dynamics in the stochastic
simulation of an interacting array of nanoparticles. The time
quantification factor is found to be the same as that derived
previously for isolated single-domain particles, up to the
linear-order in the time-step size $\Delta t$. No explicit form of
the Hamiltonian is implied in the derivation, which means that the
equivalence between the two stochastic schemes is general and
independent of many physical and simulation parameters. To
demonstrate this, we implement the TQMC scheme for the study of i)
time evolution and magnetization reversal in a square spin array,
and ii) spin wave behavior in a one-dimensional interacting array of
particles. In the case of (i), we show a close correspondence
between the TQMC and LLG results for a wide range of coupling
strength.  The equivalence remains valid even when the reversal mode
changes as $J$ is increased. In the case of (ii), the numerical
verification of the time quantification factor is provided by the
close agreement of the spin wave dispersion curves as obtained by
both TQMC and Langevin dynamical methods. The two curves also show a
very close agreement to the known theoretical dispersion relation.
Our analytical derivation and numerical studies thus justify the
applicability of the TQMC method for stochastic micromagnetic
studies in most cases where the LLG equation applies.

\section{\label{sec:levelappendix}Appendix}

\subsection{\label{sec:llg}FP coefficients for the LLG equation}

A previous study of the effect of thermal fluctuations in an
interacting spin array system based on the Langevin scheme, showed
that the inter-particle interactions do not result in any
correlations of the thermal fluctuations
\cite{chubykaloINTERACTING}. However, to the best of our knowledge,
a detailed derivation of the FP coeffcients for an interacting
particle system has not been presented. Hence we include, as an
appendix, a derivation of FP coefficients for an interacting
particle system. We extend Brown's derivation \cite{brown63} for the
FP coefficients of isolated single domain particles to obtain the FP
coefficients for the case of interacting particles.

The thermal field $\mathbf{h}(t)$ representing the thermal
fluctuations, according to Brown \cite{brown63}, has the properties
of a white noise, i.e.
\begin{equation}
\label{eq:thermalfield}
    \left<h_i^p(t)\right> = 0,\quad
    \left<h_i^p(0)h_j^q(t)\right>=2D\cdot\delta_{pq}\delta_{ij}\delta(t)
\end{equation}
where $i, j =  \{1, 2, 3\}$ denote the Cartesian coordinates
component $\{x, y, z\}$ and $p,q=\{1,\ldots,N\}$ refer to the
$p^{\mathrm{th}}$ and $q^{\mathrm{th}}$ spin in the list. Hence, if:
\begin{equation}
 \label{eq:ki}
    K_i^p \equiv \int_{t}^{t+\Delta t}h_i^p(t')dt'
\end{equation}
then
\begin{equation} \label{eq:meanKi}
    \left<K_i^p\right> = 0,\quad
    \left<K_i^pK_j^q\right> = 2D \cdot \delta_{pq} \delta_{ij} \Delta t
\end{equation}
Rewriting Eq.~(\ref{eq:llg}) in spherical coordinates, we obtain
\begin{eqnarray}
\label{eq:llgleft}
    \mbox{Left side}
      &=& \frac{d \mathbf{s}_i}{dt}
       = \frac{\partial \mathbf{s}_i}{\partial \theta_i} \frac{d \theta_i}{dt}
       + \frac{\partial \mathbf{s}_i}{\partial \varphi_i} \frac{d \varphi_i}{dt}
       = \vec{\mathbf{e}}_{\theta} \cdot \dot{\theta}_i
       + \vec{\mathbf{e}}_{\varphi} \cdot \sin \theta_i \dot{\varphi}_i
 \\
    \mbox{Right side}
      &=& \frac{\gamma_0}{\mu_0M_sV(1+\alpha^2)}
       \left(
        \mathbf{s}_i \times \frac{\partial E}{\partial \mathbf{s}_i}
        + \alpha \cdot \mathbf{s}_i \times \left( \mathbf{s}_i\times \frac{\partial E}{\partial \mathbf{s}_i}\right)
       \right)
 \nonumber\\
\label{eq:llgright}
      &=&
       \left( -h' \frac{\partial E}{\partial \theta_i}
              -g' \frac{1}{\sin \theta_i} \frac{\partial E}{\partial \varphi_i} \right) \vec{\mathbf{e}}_{\theta}
      +\left(  g' \frac{\partial E}{\partial \theta_i}
              -h' \frac{1}{\sin \theta_i} \frac{\partial E}{\partial \varphi_i} \right) \vec{\mathbf{e}}_{\varphi}
\end{eqnarray}
in which the partial differential relationships such as
$\frac{\partial E}{\partial \mathbf{s}} = \frac{\partial E}{\partial
\theta}\frac{\partial \theta}{\partial \mathbf{s}} + \frac{\partial
E}{\partial \varphi}\frac{\partial \varphi}{\partial \mathbf{s}} =
\frac{\partial E}{\partial \theta} \vec{\mathbf{e}}_{\theta} +
\frac{\partial E}{\partial \varphi} \frac{1}{\sin
\theta}\vec{\mathbf{e}}_{\varphi}$ have been used. With the
inclusion of the thermal fluctuation, additional terms will be added
into the right side as $-\frac{\gamma_0H_k}{1+\alpha^2} \left(
\mathbf{s}_i \times \mathbf{h}(t) + \alpha \cdot \mathbf{s}_i \times
\left( \mathbf{s}_i \times \mathbf{h}(t) \right) \right)$. By
considering the relation between Cartesian and spherical base
vectors:
\begin{eqnarray}
 \vec{\boldsymbol{i}} &=& \sin \theta \cos \varphi \cdot \vec{\mathbf{e}}_r
                    + \cos \theta \cos \varphi \cdot \vec{\mathbf{e}}_{\theta}
                    - \sin \varphi \cdot \vec{\mathbf{e}}_{\varphi}
 \nonumber\\
 \vec{\boldsymbol{j}} &=& \sin \theta \sin \varphi \cdot \vec{\mathbf{e}}_r
                    + \cos \theta \sin \varphi \cdot \vec{\mathbf{e}}_{\theta}
                    + \cos \varphi \cdot \vec{\mathbf{e}}_{\varphi}
 \nonumber\\
 \vec{\boldsymbol{k}} &=& \cos \theta \cdot \vec{\mathbf{e}}_r
                    - \sin \theta \cdot \vec{\mathbf{e}}_{\theta}
\end{eqnarray}
and equating Eqs. (\ref{eq:llgleft}) and (\ref{eq:llgright}), we
thus obtain $2N$ simultaneous equations as:
\begin{eqnarray}
    \frac{d\theta_i}{dt}  &=&   h' H_{\theta_i}  ' + g'\frac{1}{\sin \theta_i}H_{\varphi_i}' \nonumber\\
    \frac{d\varphi_i}{dt} &=&  -g' \frac{1}{{\sin \theta_i }}H_{\theta_i}' + h'\frac{1}{{\sin ^2 \theta_i }}H_{\varphi_i}'
  \label{eq:angularsde}
\end{eqnarray}
where
\begin{equation}
    H_{\theta_i} ' = -\frac{\partial E}{\partial \theta_i} + H_{\theta_i}
  , \quad
    H_{\varphi_i}' = -\frac{\partial E}{\partial \varphi_i} + H_{\varphi_i}
\end{equation}
$H_{\theta_i}$ and $H_{\varphi_i}$ are the contributions of
$\mathbf{h}(t)$ to the generalized forces corresponding to
$\theta_i$ and $\varphi_i$ :
\begin{eqnarray}
    (2K_uV)^{-1} H_{\theta_i}  &=& h_1^i (t)\cos \theta_i \cos \varphi_i  + h_2^i (t)\cos \theta_i \sin \varphi_i  - h_3^i (t)\sin \theta_i
  \nonumber\\
    (2K_uV)^{-1} H_{\varphi_i} &=& - h_1^i (t)\sin \theta_i \sin \varphi_i  + h_2^i (t)\sin \theta_i \cos \varphi_i
\end{eqnarray}

Eq.~(\ref{eq:angularsde}) can be expressed directly in a general
form as:
\begin{equation}
    \dot x_i^p = F_i^p(x) + \sum_{k = 1}^3 {G_{ik}^p (x)h_k^p (t)}
  \label{eq:sde}
\end{equation}
where $x$ represents the set of $2N$ variable $\{x_i^p\}$  (here $i
= \{1,2\}$ denotes angular coordinates $\{\theta,\varphi\}$ and $p =
\{1,2,\ldots,N\}$ refers to the $p^{\mathrm{th}}$ spin in the list).
To evaluate the FP coefficients $A_{x_i}$ and $B_{x_i x_j}$, we need
$\Delta x_i$ only to terms of the order $\Delta t$ for $A_{x_i}$ and
only to terms of order $(\Delta t)^{1/2}$ for $B_{x_i x_j}$. Taking
note of Eq.~(\ref{eq:ki}), $\Delta x_i$ itself is of order $(\Delta
t)^{1/2}$. Expanding $F_i^p(x)$ and $G_{ik}^p(x)$ in Taylor's series
at initial state $x_0$:
\begin{eqnarray}
    F_i^p(x) &=& F_i^p(x_0) + \sum_{q,j}F_{i,j}^{p,q}\cdot\Delta x_j^q
                + \frac{1}{2} \sum_{q,r,j,l}F_{i,jl}^{p,qr} \cdot \Delta x_j^q\Delta x_l^r + \cdots
  \nonumber\\
    G_{ik}^p(x) &=& G_{ik}^p(x_0) + \sum_{q,j}G_{ik,j}^{p,q}\cdot\Delta x_j^q
                + \frac{1}{2}\sum_{q,r,j,l}G_{ik,jl}^{p,qr} \cdot \Delta x_j^q\Delta x_l^r + \cdots
\end{eqnarray}
where, for example, $F_{i,j}^{p,q} = \partial F_i^p/\partial x_j^q$
and $G_{ik,j}^{p,q} = \partial G_{ik}^p/\partial x_j^q$. Hence by
integration of Eq.~(\ref{eq:sde}) with respect to $\Delta t$, and
truncate the terms that has order higher than $\Delta t$, we have:
\begin{equation}
    \Delta x_i^p  = F_i^p \Delta t
                    + \sum_{k} {G_{ik}^p } \int_0^{\Delta t} {h_k^p (t_1 )dt_1 }
                    + \sum_{q,j,k} {G_{ik,j}^{p,q} \int_0^{\Delta t} {\Delta x_j^q h_k^p (t_1 )dt_1 } }
\end{equation}
and in the last integral we may express $\Delta x_j^q$ to the order
of $\Delta t^{1/2}$, namely, as $\sum_{j} G_{jl}^q \int_{0}^{\Delta
t_1}h_l^q(t_2)dt_2$. Thus,
\begin{equation}
    \Delta x_i^p  = F_i^p \Delta t
                    + \sum_{k} {G_{ik}^p } \int_0^{\Delta t} {h_k^p (t_1 )dt_1 }
                    + \sum_{q,j,k,l} {G_{ik,j}^{p,q}G_{jl}^q \int_0^{\Delta t} dt_1 \int_0^{t_1} {h_k^p(t_1) h_l^q (t_2 )dt_2 } }
  \label{eq:dx}
\end{equation}
the second term is of order $\Delta t^{1/2}$, the others of order
$\Delta t$; therefore, to the first order in $\Delta t$:
\begin{equation}
    \Delta x_i^p \Delta x_j^q = \sum_{k,l} {G_{ik}^{p}G_{jl}^q
                                \int_0^{\Delta t}dt_1 \int_0^{\Delta t} {h_k^p(t_1) h_l^q (t_2 )dt_2 } }
  \label{eq:dxdy}
\end{equation}

It is easily seen that the double integral in Eq.~(\ref{eq:dx}) is
half that in Eq.~(\ref{eq:dxdy}). We now evaluate the statistical
average by considering Eq.~(\ref{eq:meanKi}) and dividing by $\Delta
t$:
\begin{eqnarray}
    A_{x_i^p} &=& \lim_{\Delta t \to 0} \frac{\left< \Delta x_i^p \right>} {\Delta t}
                = F_i^p  + D \cdot \sum_k {G_{ik,j}^{p,p} G_{jk}^{p} }
  \nonumber\\
    B_{x_i^p x_j^q} &=& \lim_{\Delta t \to 0} \frac{\left< \Delta x_i^p \Delta x_j^q\right>} {\Delta t}
                     = 2D \cdot \sum_{k}G_{ik}^{p}G_{jk}^p \cdot\delta_{pq} \delta_{ij}
 \label{eq:AnB}
\end{eqnarray}

In the present application,
\begin{eqnarray}
    F_1^p &=& -h'\frac{\partial E}{\partial \theta_p}
              -g'\frac{1}{\sin \theta_p} \frac{\partial E}{\partial \varphi_p}
  \nonumber\\
    F_2^p &=&  g'\frac{1}{\sin \theta_p}\frac{\partial E}{\partial \theta_p}
              -h'\frac{1}{\sin^2 \theta_p} \frac{\partial E}{\partial \varphi_p}
\end{eqnarray}
and
\begin{eqnarray}
\label{eq:Gii}
    (2K_uV)^{-1} G_{11}^p &=& h'\cos \theta_p \cos \varphi_p - g'\sin \varphi_p
  \nonumber\\
    (2K_uV)^{-1} G_{12}^p &=& h'\cos \theta_p \sin \varphi_p + g'\cos \varphi_p
  \nonumber\\
    (2K_uV)^{-1} G_{13}^p &=& -h'\sin \theta_p
  \\
    (2K_uV)^{-1} G_{21}^p &=& -g'\cot \theta_p \cos \varphi_p - h'\csc \theta_p \sin \varphi_p
  \nonumber\\
    (2K_uV)^{-1} G_{22}^p &=& -g'\cot \theta_p \sin \varphi_p + h'\csc \theta_p \cos \varphi_p
  \nonumber\\
    (2K_uV)^{-1} G_{23}^p &=& g'
  \nonumber
\end{eqnarray}
Partial differentiation of Eqs.~(\ref{eq:Gii}) with respect to
$\theta_p$ and $\varphi_p$ gives the formulas for the twelve
quantities $G_{ik,j}^{p,p}$ $(i,j=1,2;k=1,2,3)$. Substitution of the
values of $F_i^p$, $G_{ik}^p$ and $G_{ik,j}^{p,p}$ into
Eqs.~(\ref{eq:AnB}) gives the value of FP coefficients for LLG
dynamical equation as follows: 
\begin{eqnarray}
  \label{eq:qusildfpe}
    A^{LLG}_{\theta_i} &=& -h'\frac{\partial E}{\partial\theta_i}
                           -g'\frac{1}{\sin\theta_i}\frac{\partial E}{\partial\varphi_i}
                           +k'\cot\theta_i
  \nonumber\\
    A^{LLG}_{\varphi_i}&=& g'\frac{1}{\sin\theta_i}\frac{\partial E}{\partial\theta_i}
                           -h'\frac{1}{\sin^2\theta_i}\frac{\partial E}{\partial\varphi_i}
  \nonumber\\
    B^{LLG}_{\theta_i\theta_j}&=& 2k' \cdot\delta_{ij}
  \\
    B^{LLG}_{\varphi_i\varphi_j}&=&\frac{1}{\sin^2\theta_i} 2k' \cdot\delta_{ij}
  \nonumber\\
    B^{LLG}_{\theta_i\varphi_j}&=& B^{LLG}_{\varphi_j\theta_i} =0
  \nonumber
\end{eqnarray}
where $k' = D(h'^2+g'^2)(2K_uV)^2$ is to be determined since the
value of $D$ is still unknown. Substituting
Eqs.~(\ref{eq:qusildfpe}) into Eq.~(\ref{eq:fpe}) and taking note
that $P(\{\theta\}, \{\varphi\}, t)$ should reduce to the Boltzmann
distribution at statistical equilibrium $(\partial P /
\partial t = 0)$, one thus obtain the value of $k'$: $k' = h' /
\beta$.

\subsection{\label{sec:tqmc}FP coefficients for TQMC}

We next derive the FP Coefficients for TQMC. The Monte Carlo
algorithm starts with a random selection of the spin site. We
consider the $i^{\mathrm{th}}$ spin in the list. For a trial move
with the displacement vector to be of size $r_i$ $(r_i < R)$ and
angle $\alpha_i$ with respect to $\vec{\mathbf{e}}_{\theta}$, we
have the corresponding change with respect to $\theta_i$ and
$\varphi_i$ as \cite{chengprl}
\begin{eqnarray}
    \Delta\theta_i &=& -r_i \cos \alpha_i
                       + \frac{r_i^2}{2} \cot \theta_i \sin^2\alpha_i
                       + O\left(r_i^3\right)
  \nonumber\\
    \Delta\varphi_i &=& r_i \frac{1}{\sin \theta_i} \sin \alpha_i
                        + r_i^2 \frac{\cot\theta_i} {\sin\theta_i} \cos \alpha_i \sin \alpha_i
                        + O\left(r_i^3\right)
  \label{eq:spheretriangle}
\end{eqnarray}
The displacement probability of the size to be $r_i$ is given by
Nowak \emph{et al}. \cite{nowak} as
\begin{equation}
    p(r_i)=3 \sqrt{R^2 - r_i^2} / 2 \pi R^3
\end{equation}
and the acceptance probability for this trial move is given by the
heat bath rate as
\begin{eqnarray}
    A \left(\Delta E\right)
        &=& \frac{1}{1+\exp\left(\beta\Delta E\right)}
  \nonumber\\
        &\approx& \frac{1}{2}
          \left( 1 - \frac{1}{2}\beta
            \left(\frac{\partial E}{\partial\theta_i}\Delta\theta_i
                + \frac{\partial E}{\partial\varphi_i}\Delta\varphi_i
            \right)
          \right)
  \label{eq:six}
\end{eqnarray}
where $\Delta E$ is the energy change in the random walk step and
$\beta = (k_B T)^{-1}$. Integrating over the projected surfaces [see
Fig.~(1) in Ref.~\cite{chengprl} for a clear diagram], we obtain a
series of the required mean
\begin{eqnarray}
 \left< {\Delta \theta _i } \right> &=& \int_0^{2\pi} d\alpha _i \int_0^R (r_i dr_i) \Delta \theta _i  \cdot p(r_i) \cdot A(\Delta E) 
                                     = \frac{R^2}{20}(\cot \theta _i  - \beta \frac{\partial E}{\partial \theta _i}) + O(R^3 ) \nonumber\\
 \left< {\Delta \varphi _i } \right> &=&  - \frac{1}{{\sin ^2 \theta_i }}\frac{{R^2 }}{{20}}\beta \frac{{\partial E}}{{\partial \varphi _i }} + O(R^3 ) \nonumber\\
 \left< {\Delta \theta _i^2 } \right> &=& \frac{{R^2 }}{{20}} + O(R^4 ) \\
 \left< {\Delta \varphi _i^2 } \right> &=& \frac{1}{{\sin ^2 \theta _i }}\frac{{R^2 }}{{20}} + O(R^4 ) \nonumber\\
 \left< {\Delta \theta _i \Delta \varphi _i } \right>  &=& O(R^3 ) \nonumber
\end{eqnarray}
Let subscript $i$ ($j$) refers to the $i^{\mathrm{th}}$
($j^{\mathrm{th}}$) spin in the list and $X$, $Y$ denote either
$\theta$ or $\varphi$. One easily finds that when $i \neq j$:
$\left< \Delta X_i \Delta Y_j \right> |_{i \neq j} = 0$. This is
because in the Monte Carlo algorithm, only 1 spin site is chosen at
each Monte Carlo step. Truncating the higher order terms in the
above equations and including the probability factor of $(1/N)$ in
choosing the $i^{\mathrm{th}}$ spin from all $N$ spins, we then
obtain the FP coefficients for TQMC method as in
Eqs.~(\ref{eq:mcfpe}).


\end{document}